# The NuMI Beam at FNAL and its Use for Neutrino Cross Section Measurements

Sacha E. Kopp

*Department of Physics, University of Texas, Austin, TX 78712 USA*

**Abstract.** The Neutrinos at the Main Injector (NuMI) facility at Fermilab began operations in late 2004. NuMI will deliver an intense $\nu_\mu$ beam of variable energy (2-20 GeV). Several aspects of the design and results from runs of the MINOS experiment are reviewed. I also discuss technique to measure directly the neutrino flux using a muon flux system at the end of the NuMI line.

## INTRODUCTION

The "Neutrinos at the Main Injector" (NuMI) [1] beam line has been constructed at the Fermi National Accelerator Laboratory in Illinois, to deliver an intense $\nu_\mu$ beam to what is planned to be a variety of experiments. The first experiment, MINOS [2], will perform definitive spectrum measurements which demonstrate the effect of $\nu$ oscillations. The second, MINERvA [3], is an experiment 1 km from the NuMI target to perform neutrino cross section measurements. A third, NOvA [4], has been approved to explore the phenomenon of CP violation.

NuMI is a tertiary beam resulting from the decays of pion and kaon secondaries produced in the NuMI target. Protons of 120 GeV are fast-extracted (spill duration 8.6 µsec) from the Main Injector (MI) accelerator and bent downward by 58 mrad toward Soudan, MN, the site of the MINOS remote detector (see Figure 1). The beam line is designed to accept $4 \times 10^{13}$ protons per pulse (ppp). The design repetition rate is 0.53 Hz, giving $\sim 4 \times 10^{20}$ protons on target per year. To date, the best demonstrated intensity is $4.3 \times 10^{13}$ ppp, though in typical operations it is more like $2.5 \times 10^{13}$ ppp.

An experimental hall at approximately 1000 m from the NuMI target is home to the MINOS near detector, the MINERvA detector, and an off-axis NOvA near detector.

## BEAM LINE DESCRIPTION

The primary beam is focused onto a graphite production target of $6.4 \times 15 \times 940$ mm$^3$, segmented longitudinally into 47 fins. The beam size at the target is 1 mm. The target is water cooled via stainless steel lines at the top and bottom of each fin and is housed in an aluminum vacuum can with beryllium windows. It is electrically isolated so it can be read out as a Budal monitor [5]. After an early failure of the target due to a beam accident, the replacement target has performed quite well. Studies indicate that the existing NuMI target could withstand up to a 1 MW proton beam if the beam spot size is increased from 1 mm to 2-3 mm [6].

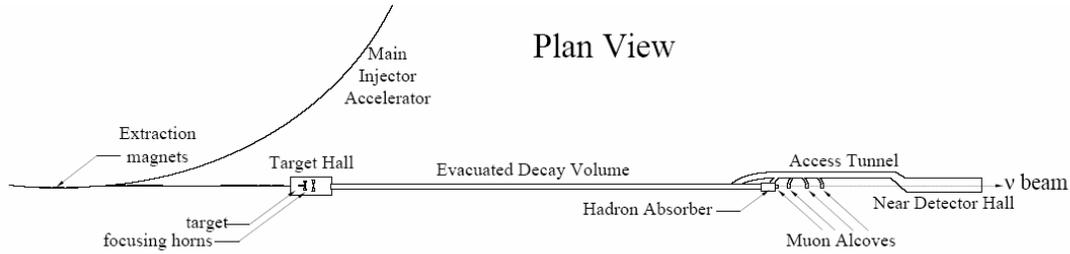

**FIGURE 1.** Plan view of the NuMI beam, indicating primary beam transport, target station, decay volume, beam absorber, muon detectors, and detector hall housing the first MINOS detector. At present, only 3 of the 4 muon alcoves are instrumented.

The particles produced in the target are focused by two magnetic "horns" [7]. The 200 kA peak current produces a maximum 30 kG toroidal field which sign- and momentum-selects the particles from the target. The relative placement of the two horns and the target optimizes the momentum focus for pions, hence the peak neutrino beam energy. The beam line at present is configured in the "Low Energy" mode with $\langle E_\nu \rangle \sim 4$ GeV for the MINOS experiment, but will switch to the "Medium Energy mode with $\langle E_\nu \rangle \sim 7$ GeV for NOvA in 2010.

To fine-tune the beam energy, the target is mounted on a rail-drive system with 2.5 m of longitudinal travel, permitting remote change of the beam energy without accessing the horns and target [8]. Such has been useful for commissioning and systematics studies in MINOS. The spectra for several target positions are shown in Figure 2.

Particles are focused by the horns into a 675 m long, 2 m diameter steel pipe evacuated to ~1 Torr. This length is approximately the decay length of a 10 GeV pion. The decay volume is surrounded by 2.5-3.5 m of concrete. At the end of the decay volume is a beam absorber which stops the remnant hadrons un-decayed in the pipe as well as the unreacted protons which traversed the NuMI target.

## SECONDARY BEAM INSTRUMENTATION

Ionization chambers are used to monitor the secondary and tertiary particle beams [9]. An array is located immediately upstream of the absorber, as well as at three muon "pits," one downstream of the absorber, one after 8 m of rock, and a third after an additional 12 m of rock. These chambers monitor the remnant hadrons at the end of the decay pipe, as well as the tertiary muons from $\pi$ and $K$ decays. When the beam is tuned to the medium energy configuration, the pointing accuracy of the muon stations can align the neutrino beam direction to approximately 50 µradians in one spill. The hadron (muon) monitor are exposed to charged particle fluxes of $10^9/cm^2/spill$ ($10^7/cm^2/spill$), and these fluxes are monitored to a precision better than 5% (1%).

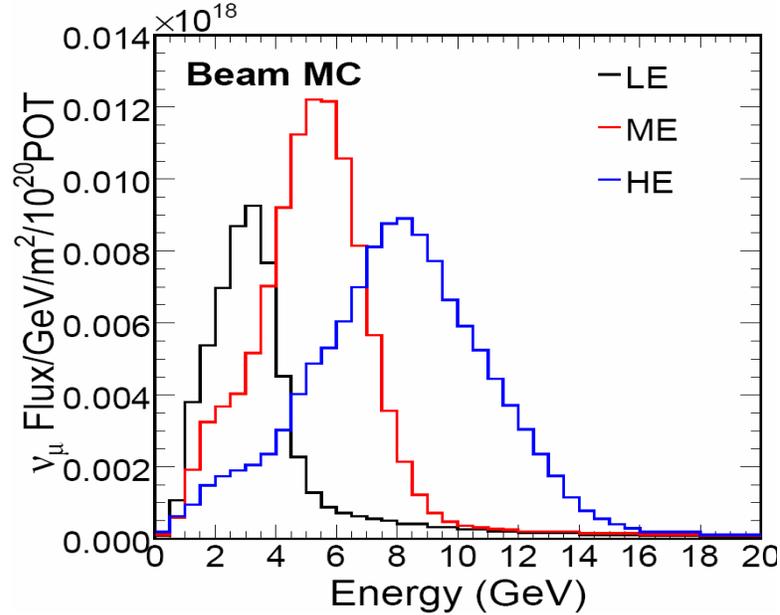

**FIGURE 2.** Neutrino energy spectra achieved at a distance of 1040 m from the NuMI target with the horns separated by 10 m and the target inside the first horn (LE), or retracted 1 m (ME) or 2.5 m (HE).

## MEASURING THE NEUTRINO FLUX

Measurement of neutrino cross sections in experiments such as MINERvA will require accurate measurement of $\sigma_\nu \equiv N_\nu/\phi_\nu$, namely both the number of neutrino interactions in the detector and the denominator, the flux of neutrinos from the beam line. Direct measurement of the flux is often accomplished by measuring the fluence of muons produced along with the neutrinos in meson decays. Cross section measurements from low-beam-power narrow-band beams such as CCFR [10] and CDHS and CHARM [11] benefited from a nearly monochromatic pencil-shaped secondary beam, which permitted insertion of instrumentation into the secondary beam as well as muon flux measurements. In a high-power horn-focused wide-band beam, one must reconstruct differential energy information of the muons only.

Following experience at BNL [12], the CERN PS [13], and IHEP [14], we propose to use the three NuMI muon alcoves to measure the differential energy distribution of the neutrino flux. Unlike these previous beam lines, however, NuMI has only 3 muon alcoves (See Figure 1), so cannot perform a differential measurement by measuring the flux of muons above each of the 4, 8, and 18 GeV/$c$ momentum thresholds of the alcoves (to be contrasted with the 7-12 muon stations in previous beam lines).

The NuMI strategy will rather be to take advantage of the flexible target/horn system which permits rapid adjustment of both the target position and the horn focusing current. By suitable variation of both the target position and horn current, one can vary in fine detail the transverse and longitudinal momenta of particles emanating from the target and sent through the beam line which contribute to the muon and the neutrino flux. The three muon stations provide three effective momentum thresholds to detect the muons (hence the hadron parents), as seen in Figure 3. Variation of the target position changes the average $\langle p_z \rangle$ of focused parent pions, while the variation of the horn current changes the transverse momentum kick given to pions, $\langle p_T \rangle$. Scanning both of these will allow us to, in effect, map out the kinematic distribution of $p_T - p_z$ of parent particles as they cross the three thresholds for reaching the three muon alcoves. In this way, a direct neutrino flux is inferred, much as in the earlier BNL, CERN PS, and IHEP beams.

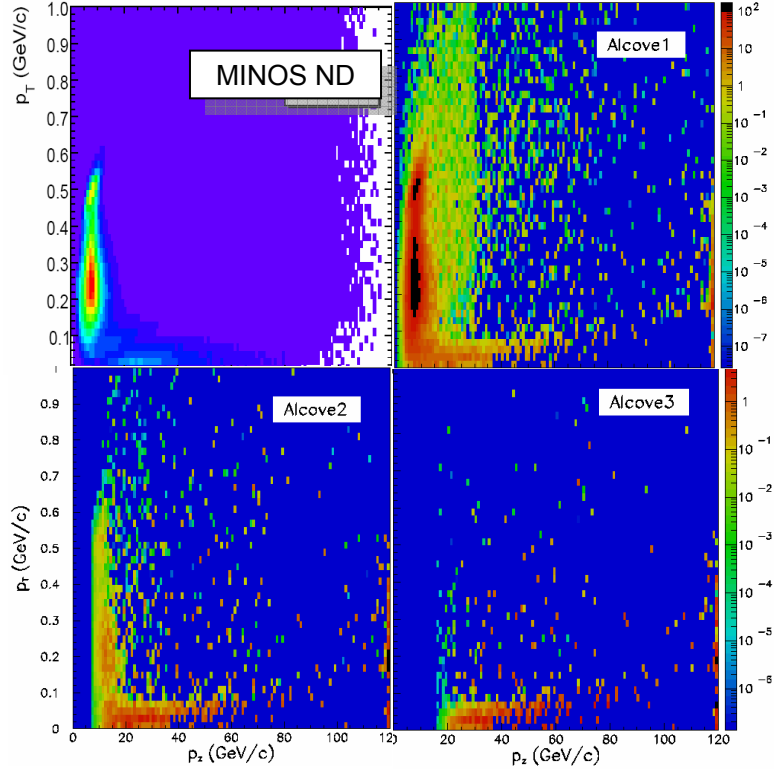

**FIGURE 3.** The transverse and longitudinal momenta of pions emanating from the NuMI target which contribute a neutrino interaction in the MINOS near detector, or a muon in Alcove 1, 2, or 3, when focused in the NuMI "LE" configuration. The horns are tuned to focus ~ 8 GeV/$c$ pions.